\documentclass[prb]{revtex4}
\begin{document}

\title{Gauge invariance and the detection of gravitational 
radiation} 
\author{David Garfinkle}
\email{garfinkl@oakland.edu}
\affiliation{Department of Physics, Oakland University, Rochester, MI 48309}

\begin{abstract}

The detection of gravitational radiation raises some subtle issues 
having to do with the coordinate invariance of general relativity.
This paper explains these issues and their resolution by using an analogy
with the Aharonov-Bohm effect of quantum mechanics. 

\end{abstract}
\maketitle

\section{Introduction}

Gravitational radiation is one of the most important predictions of 
Einstein's general theory of relativity.  It has been detected 
indirectly through its effect on the orbital period of the 
Hulse-Taylor binary pulsar\cite{taylor} but it has not yet been
directly detected.  Several gravitational radiation detectors have 
been built in an attempt to perform such a direct 
detection.\cite{barish,ogw1,ogw2,ogw3}  These
detectors are essentially large laser interferometers, and the usual
explanation of how they work goes as follows: an interferometer, by using
the interference of light beams that travel along different paths is 
very sensitive to changes in the lengths of those paths.  According to
general relativity, gravity is a distortion in the geometry of space.
When a gravitational wave passes through the detector, it changes the 
lengths of the arms of the interferometer and this change is detected
through its effect on the the relative phase of the two light rays.
At first glance, this explanation sounds simple and clear.  But on 
reflection some issues arise: one issue comes from thinking about the
usual explanation for cosmoligical redshift, which is that the expansion
of the universe causes a corresponding expansion in the wavelength of light.
Applying this concept to the interferometer, if the wavelength of the light
expands as much as the interferometer arm does, then there should be no change
in phase and therefore no detection.  Other issues arise from the fact
that general relativity, as a theory of gravity,
doesn't just give predictions for the geometry of space, but also for
the propagation of light and the motion of material objects.  In addition
to changes in the lengths of the interferometer arms then, one might
expect additional effects due to the effect of gravity on the propagation
of the light as it moves along the interferometer arms.  Furthermore,
the mirrors at each end of the arms are also subject to gravity, so one
might expect an additional effect due to motion of these mirrors 
under the effects of the gravitational wave.  Why are these additional
effects not discussed in the usual explanation of how gravitational
wave detectors work?  Are these additional effects small enough to be
negligible?  But if so, then why are they?  Are these additional
effects absent?  But again, if so, why are they absent?  

It turns out that these questions can be answered by a careful consideration
of the role of coordinate invariance in general relativity.  Though 
coordinate invariance in general relativity is not a subject easily
at the command of most physicists, it turns out that it is analogous
to gauge invariance in electrodynamics.  In fact the line of reasoning
used to resolve the issue of the properties of gravitational wave 
detectors is the same as that used to understand the role of gauge
invariance in the Aharonov-Bohm effect.  In this paper, we will first
look at the issues that come up for the Aharonov-Bohm effect and the
resolution of those issues.  We will then show how the same line of
reasoning applied to gravitational wave detectors serves to resolve the
issues raised here.

\section{Aharonov-Bohm effect}

The simplest of the Maxwell equations of electrodynamics is the statement that
the magnetic field is divergence free: 
${\vec \nabla}\cdot {\vec B}=0$.  Soon after learning this equation, we
learn a calculational trick to solve it: since the divergence of a curl
is zero, we simply introduce a vector potential $\vec A$ such that
$ {\vec B} = {\vec \nabla} \times {\vec A}$ and it then follows
that for any $\vec A$ the $\vec B$ given by this formula is automatically
divergence free.  However, we also quickly learn that the vector potential
is not unique: since the curl of a gradient is zero, it follows that for 
any scalar $\chi$, if we replace $\vec A$ by ${\vec A} + {\vec \nabla}\chi$
then the magnetic field is unchanged.  In other words, ${\vec A} + 
{\vec \nabla} \chi$ is just as good a vector potential as $\vec A$.  
This property, the invariance of $\vec B$ under the transformation
${\vec A} \to {\vec A} + {\vec \nabla }\chi $, is called gauge
invariance.  All this is not particularly disturbing.  Since $\vec A$ is a 
calculational trick rather than a physical field, there is no reason for it
to be unique.  

However, the situation changes when we consider the quantum mechanics 
of charged particles in a magnetic field.  
A standard example of quantum mechanical behavior
is the interference pattern of the two slit experiment.   
It is therefore natural to ask how this interference pattern changes when
the particles are charged and a magnetic field is applied.  
The Aharonov-Bohm effect\cite{bohm} 
consists of this case further specialized so
that the magnetic field is localized in a region between the two slits
but away from the paths that the particles would take in going from
the slits to the screen.  The wavefunction satisfies the Schrodinger
equation for a particle of charge $q$ in a magnetic field $\vec B$ which is
\begin{equation}
{\frac 1 {2m}} {{\left ( {\frac \hbar i} {\vec \nabla} - {\frac q c}
{\vec A} \right ) }^2} \psi = E \psi    
\end{equation}
where $\vec A$ is the vector potential associated with $\vec B$.  

At first sight, this equation seems crazy.  There is no such thing as 
{\it the} vector potential associated with $\vec B$.  Rather there is
a class of vector potentials, all equally good.  But if we choose a 
different $\vec A$ it appears that Schrodinger's equation will change.
Nonetheless, despite appearences Schrodinger's equation really is gauge
invariant.  It is just that whenever we make the change
${\vec A} \to {\vec A} + {\vec \nabla} \chi$ we also have to change the wave
function by
\begin{equation}
\psi \to \psi \exp (i q \chi /\hbar c)
\end{equation}  
That is, the wave function must change by a phase.  

The best way to understand this is to reverse the order of the reasoning.
We know that quantum mechanics has {\it global} phase invariance: multiply
the wave function by $\exp (i a) $ where $a$ is a real constant and 
nothing physical changes.  However, it turns out that the quantum 
mechanics of charged particles also has {\it local} phase invariance:
multiply the wave function by $\exp (i f ({\vec x}))$ and nothing 
physical changes.  In order for nothing to change under a local change of
phase, the theory must contain a vector field $\vec A$ 
coupled to the wave function
in a particular way and the vector field must change by  
\begin{equation}
{\vec A} \to {\vec A} + {\frac {\hbar c} q} {\vec \nabla} f
\end{equation}
whenever $\psi $ is multiplied by  $\exp (i f ({\vec x}))$.  The gauge
invariance of electrodynamics is then not the byproduct of a particular
mathematical trick to calculate the magnetic field.  Instead gauge
invariance is a rather deep property having to do with the local
phase invariance of quantum mechanics.   

Now we calculate the effect of the magnetic field on the interference 
pattern.  The particles come from a source at point $a$, go through
either slit 1 or slit 2, and arrive at a point $b$ on the screen.  By
superposition, the wavefunction at point $b$ takes the form
$\psi = {\psi _1}+{\psi _2}$ where $\psi _1$ is the value that the wave
function would have if only slit 1 were open, and correspondingly for
$\psi _2$.  Now, let $\psi _{10}$ be the value that $\psi _1$ would
have if the magnetic field were turned off, and correspondingly for
$\psi _{20}$.  For all points $\vec x$ on the far side of the slits, 
define the function $W_1$ by
\begin{equation}
{W_1}({\vec x})={\int _{C_1}} {\vec A} \cdot d {\vec l} 
\label{Wdef}
\end{equation}
Here the curve $C_1$ is essentially two straight lines: 
one from  $a$ to slit 1
and the other from slit 1 to $\vec x$, though we round off the curve at
slit 1 to make it smooth.  Here $\vec x$ is any point on the far side of the 
slits, though in particular we will apply equation (\ref{Wdef}) to points
$b$ on the screen. 
Correspondingly define $W_2$.  The point of this somewhat clumsy looking
definition is that since $W_1$ is the line integral of $\vec A$ it
follows that ${\vec \nabla}{W_1} ={\vec A}$.  Then using $\chi = -{W_1}$ in the
gauge transformation ${\vec A} \to {\vec A} + {\vec \nabla}\chi$
transforms $\vec A$ to zero in a region that includes points $a$ and
$b$ and slit 1.  It then follows from gauge invariance that
\begin{equation}
{\psi _1} = {\psi _{10}} \exp (i q {W_1}/\hbar c)
\end{equation}   
and correspondingly for $\psi _2$.  Let $\Delta \phi$ be the change 
in phase difference between $\psi _2$ and $\psi _1$ due
to the magnetic field.  Then it
follows that
\begin{equation}
\Delta \phi = {\frac q {\hbar c}} ({W_2} - {W_1}) =  
{\frac q {\hbar c}} {\int _C} {\vec A} \cdot d {\vec l}
=  {\frac q {\hbar c}} {\int _S} {\vec B} \cdot d {\vec a}
\end{equation}
Here the closed curve $C$ is the curve $C_2$ from $a$ to $b$ followed by the
curve $C_1$ backwards from $b$ to $a$.  The surface $S$ is one whose 
boundary is the curve $C$.  Note that ${\int _S} {\vec B} \cdot d {\vec a}$
is simply the magnetic flux $\Phi$ through that surface.  Thus we have
\begin{equation}
\Delta \phi = {\frac q {\hbar c}} \Phi
\label{giabphase}
\end{equation}

In this problem the vector potential $\vec A$ and its 
gauge invariance properties
were sources of both conceptual confusion and some helpful calculation
techniques.  Note however that $\vec A$ is completely 
absent in the final result.  Instead, the final result relates 
the physically measured quantity, the phase difference, to a gauge
invariant quantity, the magnetic flux.  For conceptual clarity, this 
is the sort of formula that we would like in a theory
that has gauge.
Though we may use a particular gauge to do the calculation; things are
much more clear if the final result is explicitly gauge invariant.  It is
this sort of result that we will derive for gravitational wave 
detection in the next section. 

\section{Gravitational wave detectors}

General relativity is the theory of the metric $g_{\mu \nu}$ of 
curved spacetime.  In a strong gravitational field, including those
where the gravitational waves of interest are produced, the field 
equations of general relativity are highly nonlinear and somewhat 
complicated.  However, once the gravitational waves get to the
detector, they are quite weak.  The metric can then be expressed
as 
\begin{equation}
{g_{\mu \nu}} = {\eta _{\mu \nu}} + {h_{\mu \nu}}
\end{equation}
Here $h_{\mu \nu}$ is small and $\eta _{\mu \nu}$ is the flat metric of
special relativity.  That is in the usual coordinates 
$(t,x,y,z)$ we have
${\eta _{tt}} = -1 $ and
${\eta _{xx}}={\eta _{yy}}={\eta _{zz}}=1$ and all other components vanish.
Since $h_{\mu \nu}$ is small, we can write all equations in first order
perturbation theory.  Thus, we can treat gravity as the theory of
a tensor field $h_{\mu \nu}$ in special relativity in much the same
way that electrodynamics is the theory of a vector field $A_\mu$ (the 
four-vector version of the vector potential).  We use the relativity 
conventions that greek indicies stand for spacetime components
($t,x,y,z$), latin
indicies for spatial components ($x,y,z$)
that repeated upper and lower indicies are summed over and that units are 
chosen so that the speed of light is equal to one.
We now consider the
motion of material particles and light rays.  We usually think of a 
trajectory as giving the spatial coordinates $\vec x$ as a function
of time.  For our purposes, it will be helpful to instead give all
four coordinates $x^\alpha$ as functions of a parameter $\lambda$ called
the affine parameter.  For material particles this affine parameter
will be the proper time, that is the time elapsed on a clock carried along
that particle's trajectory.  For a light ray, the affine parameter
will be the phase.  Let an overdot denote derivative with respect
to $\lambda$ and define ${u^\alpha}={{\dot x}^\alpha}$.  Then the
equation of motion is 
\begin{equation}
{{\dot u}^\alpha} + {\Gamma ^\alpha _{\beta \gamma}}
{u^\beta}{u^\gamma} = 0
\end{equation}
Here the Christoffel symbols 
$ \Gamma ^\alpha _{\beta \gamma}$ are given by
\begin{equation}
{\Gamma ^\alpha _{\beta \gamma}} = {\textstyle {\frac 1 2}}
{\eta ^{\alpha \delta}} \left ( {\partial _\beta}{h_{\gamma \delta}}
+ {\partial _\gamma}{h_{\beta \delta}}-{\partial _\delta}
{h_{\beta \gamma}}\right ) 
\label{Chr}
\end{equation} 
where $\partial _\alpha$ is an abreviation for the partial derivative 
with respect to $x^\alpha$.  

Since general relativity is invariant under general coordinate transformations,
some form of this invariance must be inherited by the weak field
form of the theory.  A (small) coordinate change takes the form 
${x^\alpha} \to {x^\alpha} - {\xi ^\alpha} $ 
Where the $\xi ^\alpha$ are functions of the $x^\alpha$.  Under this 
coordinate change the $h_{\alpha \beta} $ transform as
\begin{equation}
{h_{\alpha \beta}} \to {h_{\alpha \beta}} + {\partial _\alpha} {\xi _\beta}
+ {\partial _\beta} {\xi _\alpha} 
\label{diffeo}
\end{equation} 
This is the gauge transformation of weak field gravity.  Note that the
Christoffel symbols are {\it not} gauge invariant and therefore the equations
of motion are {\it not} gauge invariant.  Thus, in particular, the question
``what is the trajectory of a material particle or a light ray?'' 
is not a question with a gauge
invariant answer.  We will thus have to be a bit careful about what sort
of questions we ask.    

What then is gauge invariant in weak field gravity?  The answer is a 
quantity called the Riemann tensor.  This can be expressed in terms
of the Christoffel symbols as
\begin{equation}
{R_{\alpha \beta \gamma \delta}} ={\eta _{\delta \epsilon}}
\left ( {\partial _\beta}{\Gamma ^\epsilon _{\alpha \gamma}}
- {\partial _\alpha}{\Gamma ^\epsilon _{\beta \gamma}}\right )
\end{equation}
or in terms of $h_{\alpha \beta}$ as
\begin{equation}
{R_{\alpha \beta \gamma \delta}} = {\textstyle {\frac 1 2}}
\left ( {\partial _\beta}{\partial _\gamma}{h_{\alpha \delta}}
+ {\partial _\alpha}{\partial _\delta}{h_{\beta \gamma}}
- {\partial _\beta}{\partial _\delta}{h_{\alpha \gamma}}
- {\partial _\alpha}{\partial _\gamma}{h_{\beta \delta}}\right )
\label{Riem}
\end{equation}
Using equation (\ref{diffeo}) in equation (\ref{Riem}) shows that 
the Riemann tensor does not change under a gauge transformation.
The Riemann tensor has a straightforward physical interpretation in
terms of tidal force.  Recall that in a non-uniform gravitational field, 
objects at different positions acquire different accelerations.  It is 
this effect, from the gravitational field of the moon, that leads to 
ocean tides.  For two (slowly moving) objects with separation 
$\Delta {\vec s}$, their relative acceleration $\Delta {\vec a}$
is given by
\begin{equation}
\Delta {a^i}=- {R_{tkti}}\Delta {s^k}
\label{geodev}
\end{equation}

Recall that the relative phase of the light rays in the interferometer is
a physically measured quantity.  Our goal, in analogy with the previous
section of the paper is to find a formula 
that expresses that quantity in terms of the
Riemann tensor.   That formula will then be manifestly gauge invariant.
Though we may end up introducing a particular gauge to do the calculation,
once we have the result we will no longer need to worry about issues of
gauge.   

The ability to make gauge transformations is the ability to choose a 
gauge that makes the calculations convenient.  This is a familiar
procedure for electrodynamics.  For our purposes, a convenient gauge
is the so called ``radiation gauge'' in which 
(among other properties of the gauge) $h_{\mu \nu}$ has only
spatial components, in other words where $h_{tt}$ and $h_{ti}$ 
vanish.  For gravitational radiation, $h_{\mu \nu}$ can always be
put in radiation gauge.  From equation (\ref{Chr}) it follows that 
$\Gamma ^i _{tt}$ vanishes in radiation gauge.  Therefore, a solution
to the equation of motion for material particles is that the spatial
coordinates are constant (and therefore that ${u^i}=0$).  
Thus the mirrors of the interferometer do not
change their spatial coordinates under the influence of the gravitational
wave.  Note that this is not the same as saying that 
``the mirrors do not move.''  It is only a statement about the coordinates
of the mirrors in a particular gauge.  From equation (\ref{Riem}) 
it follows that
(in this gauge)
\begin{equation}
{R_{tkti}} = - {\textstyle {\frac 1 2}} {\partial _t}{\partial _t}{h_{ki}}
\end{equation}    

We now treat the propagation of light in the interferometer.  Though the 
interferometer is large (arm lengths of 4 km for LIGO), it is small 
compared to the wavelength of the gravitational radiation that it is
designed to detect (the peak sensitivity of LIGO is at about 200 Hz 
corresponding
to a wavelength of about 1500 km).\cite{barish}  
This means that during a single trip of a light
ray in an arm
of the interferometer, the components of $h_{\mu \nu}$ can be regarded
as constant: having no dependence on space or time.  (Note that this is
not true of the proposed space based gravitational wave detector LISA, 
which is much larger and for which the analysis of this paper would
be more complicated).  Therefore for the purposes of calculating 
the properties of a single light ray trip, the $g_{\mu \nu}$ are
constant with ${g_{tt}}=-1$ and ${g_{ti}}=0$.  That is, the metric
is just the usual metric of special relativity with $t$ as the usual
time.  However the $x^i$ are not the usual spatial coordinates because
the $g_{ij}$, though constant, no longer have their flat space values of
$1$ for the diagonal components and $0$ for the off-diagonal ones.  Instead,
the usual spatial coordinates are some linear combination of the $x^i$.
Thus, the only tools that we will need to find the phase difference
of the two light rays are the ordinary rules of light propagation
of special relativity.  These rules say that for light of 
angular frequency
$\omega$ the phase of the wave is $\phi = \omega (t-l)$ where $t$
is the special relativity time (and because of the radiation gauge, also
our coordinate time) and where $l$ is the spatial distance in the 
direction along which the light propagates.  The phase difference
between the two light rays is then equal to $\omega$ multiplied by
the difference in their travel times and is therefore also equal to
$\omega$ multiplied by the difference in their travel distances.   

Choose the origin of 
coordinates to be the position of the beam splitter and the $x$ and $y$
axes to contain the two mirrors at the ends of the interferometer arms.
Then the difference in phase between the two light rays when they
recombine at the beam splitter is
\begin{equation}
\Delta \phi = \omega \Delta x (2 + {h_{xx}}) - \omega \Delta y 
(2 + {h_{yy}})  
\label{delphi}
\end{equation}
Here $\Delta x$ is the $x$ coordinate position of the mirror on the 
$x$ axis, and correspondingly for $\Delta y$.  The first term in each 
parenthesis is much larger than the second.  In the absence of 
gravitational radiation, this term is just the usual expression for
the phase in an interferometer having to do with the difference  in the
lengths of the arms.  However, this term is time independent; so we
can get rid of it by taking a time derivative.  In fact we will
take two time derivates of equation (\ref{delphi}) yielding
\begin{equation}
{\frac {{d^2}{\Delta \phi}} {d {t^2}}} = \omega  (
\Delta x {\partial _t}{\partial _t} {h_{xx}} - \Delta y {\partial _t}
{\partial _t}{h_{yy}}  )
\end{equation}
Since terms involving $h_{\mu \nu}$ are first order, we can replace
$\Delta x$ and $\Delta y$ by their zeroth order expressions:
$\Delta x = \Delta y = L$ where we are assuming (approximately) 
equal lengths $L$ for the interferometer arms.  Since we are in
radiation gauge, we can replace second derivatives of the metric
with the Riemann tensor.  We thus have
\begin{equation}
{\frac {{d^2}{\Delta \phi}} {d {t^2}}} = 2 \omega L 
( {R_{tyty}}-{R_{txtx}})
\label{giligophase}
\end{equation}
Equation (\ref{giligophase}) is the main result of this paper.  
It expresses a physically
measurable quantity, the second time derivative of the phase difference,
in terms of a gauge invariant quantity, the Riemann tensor.  It is thus
the analog for gravitational radiation of equation (\ref{giabphase}) 
for the Aharonov-Bohm effect.  

We are now in a position to address the issues raised in the introduction.
The cosmological redshift is due to the expansion of the universe
between the time the light is emitted and the time it is absorbed.  
We have neglected this sort of effect in the interferometer by
treating the $h_{\mu \nu}$ as constants between the time the beam 
is split and the time it recombines.  A more exact treatment not making
this approximation does yield an additional effect.  However, this
additional effect is smaller by a factor of $L/{\tilde \lambda}$ 
(where $L$ is the interferometer arm length and $\tilde \lambda$ is 
the wavelength of the gravitational radiation) than the effect we have
calculated.  As for the effect of gravitational radiation on the motion of
the mirrors, though the mirrors are at constant spatial coordinates, their
separation is not constant.  The acceleration of their separation is
given by equation (\ref{geodev}).  Whether one regards the 
change in separation
as a consequence of motion of the mirrors or the expansion (and contraction)
of space is a matter of semantics.  What are not matters of semantics are
observable quantities like the phase difference between the two light beams.
   
The examples of this section and the previous one illustrate an important
principle of physics that recurs throughout both quantum mechanics and
general relativity: a physical theory is required to answer only those 
questions that correspond to a measurement.  When the theory in question
is a gauge theory, the answer about the result of a measurement is
gauge invariant.  However, for convenience the calculations are often done
in a particular gauge.  Often the results are simply presented in that 
gauge, and if one forgets that a particular gauge choice has been made
it seems as if a physical quantity (the result of the measurement) has
been equated to a quantity without an unambiguous meaning.  
Another way to put this issue is that it is important to remember that
the results of a newer theory (general relativity) cannot be ``shoehorned''
into the categories of the old theory (Newtonian mechanics) even though 
it is tempting to try to do so.  What is physically meaningful in one
theory may be mere gauge in the other.  
This temptation is particularly dangerous
when the results of a calculation in a particular gauge seem to have a
nice physical interpretation using the old theory categories.   
One might then regard the calculation as the`` physical'' explanation
of the phenomenon and become confused when a calculation in a different
gauge seems to point to a different ``physical'' explanation.  
To allay 
this confusion, it is sometime helpful to present the results in a way
that is manifestly gauge invariant by expressing the result of the
measurement in terms of gauge invariant quantities.  This is what
has been done in this paper for gravitational radiation detectors.   

\section{Acknowledgements}

I would like to thank Alberto Rojo for posing the question that this paper
answers.
This work was partially supported by
NSF grant PHY-0456655 to Oakland University.


\begin{thebibliography}{2}

\bibitem{taylor}
J. Taylor and J. Weisberg, 
{\it A New Test of General Relativity: Gravitational Radiation and
the Binary Pulsar PSR 1913+16}, Astrophys. J. {\bf 253}, 908-920 (1982)

\bibitem{barish}
B. Barish and R. Weiss, 
{\it LIGO and the Detection of Gravitational Waves}, 
Phys. Today {\bf 52}, 44-50 (1999)

\bibitem{ogw1}
B. Willke et al., 
{\it The GEO 600 gravitational wave detector}, 
Class. Quant. Grav. {\bf 19}, 1377-1388 (2002)

\bibitem{ogw2}
F. Acernese et al., 
{\it The present status of the VIRGO Central Interferometer}, 
Class. Quant. Grav. {\bf 19}, 1421-1428 (2002)

\bibitem{ogw3}
H. Tagoshi et al., 
{\it First search for gravitational waves from inspiraling compact binaries
using TAMA300 data}, Phys. Rev. {\bf D63}, 062001-(1-5) (2001) 

\bibitem{bohm}
Y. Aharanov and D. Bohm, 
{\it Significance of Electromagnetic Potentials in the Quantum Theory}, 
Phys. Rev. {\bf 115}, 485-491 (1959)


\end{thebibliography}
\end{document}